\documentstyle[12pt] {article}
\newcommand{\beq}{\begin{equation}}
\newcommand{\eeq}{\end{equation}}
\newcommand{\beqa}{\begin{eqnarray}}
\newcommand{\eeqa}{\end{eqnarray}}
\newcommand{\ba}{\begin{array}}
\newcommand{\ea}{\end{array}}

\begin{document}

\begin{flushright}
To be published\\
Nuovo Cimento A
\end{flushright}

\begin{center}
{\large \bf The Onset of Chaos with a\\
Quadrupole--Quadrupole Interaction $^{(*)}$}
\end{center}
\vspace{0.5 cm}

\begin{center}
{\bf V.R. Manfredi}  \\
Dipartimento di Fisica ``G. Galilei" dell'Universit\`a
di Padova, \\
INFN, Sezione di Padova, \\
Via Marzolo 8, I 35131 Padova, Italy $^{(+)}$ \\
Interdisciplinary Laboratory, SISSA, Trieste \\
\end{center}

\vskip 0.5 truecm

\begin{center}
{\bf L. Salasnich}  \\
Dipartimento di Fisica dell'Universit\`a di Firenze, \\
INFN, Sezione di Firenze, \\
Largo E. Fermi 2, I 50125 Firenze, Italy \\
\end{center}

\vskip 1. truecm
{\begin{center}
Preprint DFPD/93/TH/73\\
\end{center}}
\vskip 0.7 truecm
-------------------------- \\
$^{(*)}$ ~This work has been partially supported by the Ministero \\
dell'Universit\`a e della Ricerca Scientifica e Tecnologica (MURST).\\
$^{(+)}$ Permanent address. E--Mail: manfredi@padova.infn.it
\newpage

\begin{center}
{\bf Abstract}
\end{center}
\vskip 0.5 truecm
\par
The transition from order to chaos in atomic nuclei has been studied
analytically and numerically using a quadrupole--quadrupole residual
interaction. This interaction leads to chaotic behaviour,
but the critical energy $E_C\simeq 12.6$ MeV, corresponding
to the onset of chaos, is higher than that of the experimental one.

\newpage

\par
{\bf 1. Introduction}
\vskip 0.5 truecm
\par
As is well known, in many--body systems states of different
"complexity\footnote{In this paper the terms "complexity" and
"hierarchy" have been used in a general sense; for precise definitions,
see [17].}" generally coexist [1,2,3].
In particular, in atomic nuclei the "hierarchy"
of complexity can be represented by a simple picture.
In zero order approximation the relevant elementary excitations
(single--particle, rotations, vibrations) may regarded as
independent modes and then
the interaction between these elementary modes can be considered [4].
At the end of the chain of complexity there are the so called "stochastic"
or "chaotic states", whereas a single mode can be considered
"regular".
\par
The experimental data of nuclear spectroscopy (see, e.g. references
[5,6]) suggest regular states near the ground state ($0$--$3$ MeV) and
chaotic states near the neutron emission threshold ($6$--$8$ MeV).
On the other hand, in the framework of the so called "rotational
damping", the authors of references [7] have shown that the surface delta
interaction produces chaotic behaviour while a pairing plus quadrupole
force is unable to do so. The aim of this paper is to analyze
{\it in detail} the quadrupole--quadrupole residual interaction
showing that it is able to reproduce the transition order--chaos
in the atomic nuclei, but at an energy ($\sim 12.6$ MeV)
higher that that of the experimental one ($\sim 8$ MeV).
\par
The structure of the article is as follows: in section 2 we present a
schematic nuclear model that describes two nucleons in a mean--field
interacting with a separable quadrupole--quadrupole residual
interaction; in section 3 we apply the curvature of
potential energy criterion to calculate the energy $E_C$
of the onset of chaos and
in section 4 we plot the Poincar\`e sections of the model for bound
energies. Finally in section 5 we diagonalize the quantum hamiltonian
and calculate the distribution $P(S)$ of spacings between adjacent
levels.

\vskip 0.5 truecm
\par
{\bf 2. The schematic quadrupole--quadrupole interaction}
\vskip 0.5 truecm
\par
In our model nucleons move independently in an oscillator
potential under the influence of a separable
quadrupole--quadrupole effective residual interaction. The hamiltonian
of the system is given by [8]:
$$
H=\sum_{i=1}^A {p_i^2\over 2m}+V_0\beta^2(A-1)\sum_{i=1}^A r_i^2 -
$$
\beq
-{16\pi V_0\over 15}\beta^4
\sum_{i<j}^A \sum_{m=-2}^{2} (-1)^m \{ r_i^2 Y_{2m}(\theta_i,\phi_i)\}
\{ r_j^2 Y_{2-m}(\theta_j,\phi_j) \},
\eeq
where $R_0=1/\beta$ is the range of the interaction and $A$ the number
of nucleons. We consider only 2 nucleons and
suppose that they move in one dimension,
so that $Y_{lm}=\sqrt{({2l+1\over 4\pi})}\delta_{m,0}$, and we have:
\beq
H={1\over 2m}(p_1^2+p_2^2)+V_0\beta^2 (r_1^2+r_2^2)
-{4 \over 3} V_0 \beta^4 r_1^2 r_2^2.
\eeq
To simplify the problem we perform the transformation:
\beq
p_k \to \sqrt{1\over 2m\hbar \omega} p_k,
\;\;\;\;
q_k \to \sqrt{m\omega \over 2\hbar} q_k
\eeq
where $m\omega^2/2=V_0/R_0^2$. So the hamiltonian becomes:
\beq
H=\epsilon(p_1+p_2+q_1+q_2)-\chi q_1^2 q_2^2,
\eeq
with $\epsilon =\hbar \omega$ and $\chi =8/3 (\hbar^2/ mR_0^2)$.
\par
In nuclei $\epsilon=\hbar\omega\simeq 41\; A^{-1/3}$ MeV is the
single particle energy for the harmonic oscillator potential [8] and
$R_0=1/\beta\simeq 1.2 A^{1/3}$ fm is the nuclear radius, and so we
have $\chi \simeq 74 A^{-2/3}$ MeV.

\vskip 0.5 truecm

{\bf 3. The onset of chaos with the curvature criterion}
\vskip 0.5 truecm
\par
We apply the criterion of curvature [9] to the
hamiltonian (4) in order to calculate the critical energy of the transition
to chaos, which will be indicated by $E_C$.
The criterion of curvature is based on the estimation of the rate of separation
of neighbouring trajectories in the phase space. To calculate
the time evolution of the dynamical system with hamiltonian:
\beq
H={1\over 2m}(p_1+p_2)+V(r_1 ,r_2 )
\eeq
the following equations have to be solved:
\beq
{d\over dt} {\vec r} = {\partial H \over \partial {\vec p}} \quad , \quad
{d\over dt} {\vec p} = -{\partial H \over \partial {\vec r}},
\eeq
where ${\vec r}=(r_1,r_2)$ and ${\vec p}=(p_1,p_2)$.
The deviations from the two initially neighbouring trajectories
$({\hat r}(t),{\hat p}(t))$ are given by:
\beq
\delta {\vec r}(t)={\hat {\vec r}}(t)-{\vec r}(t),
\quad \delta {\vec p}(t)={\hat {\vec p}}(t)-{\vec p}(t)
\eeq
and the linearized equations of motion for the deviations are:
\beq
{d\over dt}\delta {\vec p}(t)=M^{-1} \delta {\vec p}(t), \quad
{d\over dt}\delta {\vec r}(t)=-S(t) \delta {\vec r}(t)
\eeq
where $M_{ij}^{-1}=\delta_{ij}m^{-1}$, and:
\beq
S_{ij}(t)={\partial^2 V\over
\partial r_{i}\partial r_{j}}|_{{\vec r}={\vec r}(t)}.
\eeq
The stability of the dynamical system is then determined by the eigenvalues
of the $4 \times 4$ matrix:
\beq
\Gamma (t)= {\left(\matrix {0 & M^{-1} \cr -S(t) & 0 \cr } \right) }.
\eeq
If at least one of the eigenvalues $\lambda_i (t)$ is real, then the separation
of the trajectories grows exponentially, and the motion is unstable.
Imaginary eigenvalues correspond to stable motion. To diagonalize the
matrix $\Gamma (t)$, we must first solve the equations of motion of the
differences (8). The problem can be significantly simplified by assuming
that the time dependence can be eliminated, i.e. $\Gamma ({\vec r}(t))=
\Gamma ({\vec r})$.
The eigenvalues then are:
\beq
\lambda_{1,2,3,4}=\pm[-b\pm\sqrt{b^2-4c}]^{1\over 2},
\eeq
where:
\beq
b=m^{-1}[{\partial^2 V\over \partial r_1^2}+
{\partial^2 V\over \partial r_2^2}],
\eeq
\beq
c=m^2[{\partial^2 V\over \partial r_1^2}
{\partial^2 V\over \partial r_2^2}-
({\partial^2 V\over \partial r_1 \partial r_2})^2].
\eeq
Now, if $b>0$ then with $c\geq 0$ the eigenvalues are purely imaginary and the
motion is stable, meanwhile with $c<0$ the pair of eigenvalues become real,
and this leads to exponential separation of neighbouring trajectories, i.e.
chaotic motion.
The parameter $c$ has the same sign as the Gaussian curvature
$K(r_1 ,r_2 )$ of the potential--energy surface:
\beq
K(r_1 ,r_2 )={ {\partial^2 V\over \partial r_1^2}
{\partial^2 V\over \partial r_2^2}-
({\partial^2 V\over \partial r_1 \partial r_2})^2
\over [1+({\partial^2 V\over \partial r_1^2})^2+
({\partial^2 V\over \partial r_2^2})^2]^2 }.
\eeq
Let us now return to our nuclear problem. The  potential energy is:
\beq
V(r_1,r_2)=\epsilon (r_1^2+r_2^2)
-\chi r_1^2 r_2^2;
\eeq
it has one minimum for $r_1=r_2=0$ with $V=0$ MeV, and four saddle
points for $r_1=\pm \sqrt{\epsilon\over \chi}$,
$r_2=\pm \sqrt{\epsilon\over \chi}$, and
$r_1=\pm \sqrt{\epsilon \over \chi}$,
$r_2=\mp \sqrt{\epsilon\over \chi}$, with
$V={\epsilon^2\over \chi}\simeq 23.3$ MeV.
These are the points for which the curvature criterion is exact ($\nabla
H =0$); the origin is a stable elliptic point with
$\lambda_{1,2,3,4}=\pm \sqrt(2) i$ and the four saddle points are
unstable hyperbolic points with
$\lambda_{1,2}=\pm 2$ and $\lambda_{3,4}=\pm 2 i$.
\par
The equipotentials (curves of constant potential $V$) are shown in
figure 1, and we see that {\it unbounded motion} occurs if
$E>V(\sqrt{\epsilon\over \chi},\sqrt{\epsilon\over \chi})=
{\epsilon^2\over \chi}$ and {\it bounded motion} for $E<{\epsilon^2\over
\chi}$. We concentrate here upon the region $0\leq E \leq
{\epsilon^2\over \chi}$.
\par
At low positive energies, the motion near the minimum of the potential
energy, where the curvature is positive, is periodic or quasi--periodic and is
separated from the region of instability by a line of zero curvature.
If the energy is increased, the system will, for certain initial
conditions, be in a region of negative curvature where the motion is chaotic.
In accordance with this scenario, the energy of order$\to$chaos transition
$E_C$ is equal to the minimum value of the line of zero gaussian
curvature $K(r_1 ,r_2 )$ of the potential--energy surface of the system.
For our potential, the gaussian curvature vanishes at the points
that satisfy the equation:
\beq
4\epsilon (1-{\chi\over \epsilon} r_1^2 -{\chi \over \epsilon}
r_2^2 -3 {\chi^2 \over \epsilon^2} r_1^2 r_2^2) =0.
\eeq
The energy on the zero--curvature line is determined by the expression:
\beq
V(K=0,r_2 )={\epsilon \chi^2 r_2^4 \over 3\chi^2 r_2^2 +\epsilon \chi}-
{2\epsilon^2 \chi r_2^2 \over 3 \chi^2 r_2^2 +\epsilon \chi}
+{\chi^3 \over 3\chi^2 r_2^2 +\epsilon \chi}+\epsilon r_2^2,
\eeq
(see also fig. 2). It is easy to show that the minimal energy on the
zero--curvature line is given by:
\beq
V_{min}(K=0,\bar{r_1})={5\epsilon^2\over 9\chi},
\eeq
and occurs at $\bar{r_1}=\pm {\sqrt{2}\over 2\beta}$.
This is the critical energy of the transition to chaos $E_C$
of the model, and we have $E_C\simeq 12.62$ MeV. We observe that the
energy of the onset of chaos ($E_C\simeq 12.6$ MeV) is much higher
than that required for the atomic nuclei one
(chaotic behaviour $\simeq$ 8 MeV) [5,6].
\par
In figure 3 we show the rate of regular points of configuration space
obtained with the curvature criterion; the transition to chaos seems to
be quite smooth.

\vskip 0.5 truecm
{\bf 4. Classical calculations}
\vskip 0.5 truecm
\par
The curvature criterion is only able to
characterize the local behaviour of the system (for example local
instability) and gives a rough signature of the global
properties (e.g. global instability) [10]. As is well known,
global properties are more appropriately studied using
Poincar\`e sections [12,13].
\par
The classical equations of motion obtained from the hamiltonian (4) are:
$$
{\dot r_1}=2\epsilon p_1,
\;\;\;\; {\dot r_2}=2\epsilon p_2
$$
\beq
{\dot p_1}=-2\epsilon r_1 +2\chi r_1 r_2^2,
\;\;\;\;
{\dot p_2}=-2\epsilon r_2 +2\chi r_1^2 r_2.
\eeq
We used a fourth order Runge--Kutta method [11] to compute the classical
trajectories. Poincar\`e sections [12,13] have been produced for a
variety of energies, showing the transition from almost totally regular
motion at low energies to almost totally irregular motion at high energies.
Conservation of energy restricts any trajectory in four--dimensional
phase space to a three--dimensional energy shell. At a particular
energy, therefore, the restriction $r_1=0$ defines a two--dimensional
surface in phase space. Each time a particular trajectory passes through
the surface, i.e. each time it crosses the $r_2$ axis, a point is
plotted at the position of intersection $(r_2,p_2)$. We employ a
first--order interpolation process to reduce inaccuracies due to the use
of a finite step length [13].
\par
Regular regions on the surface of section plots are characterised by
sets of invariant intersection points. The surface of section pictures
(fig. 4) show that motion is almost wholly regular for $E=15$ MeV.
However, for $E=21$ MeV we see evidence of bifurcation: one can count at
least 6 new elliptic points and some points are distributed irregularly.
For $E=23$ MeV the number of irregularly distributed points increases
but many regular regions remain.
\par
These numerical results are in good agreement with the smooth transition
to chaos predicted by the curvature criterion.

\vskip 0.5 truecm
{\bf 5. Quantum calculations}
\vskip 0.5 truecm
\par
In quantum mechanics one cannot apply classical concepts and methods
directly since the notion of trajectory is absent. Nevertheless, many
efforts have been made to establish the features of quantum systems
which reflect the qualitative difference in the behaviour of their
classical counterparts [2]. Many schematic models [2,3] have shown that
this difference reveals itself in the properties of fluctuations in
eigenvalue sequences. The spectral statistics for the systems with
underlying chaotic behaviour agree with the predictions of the random matrix
theory. By contrast, quantum analogs of classically integrable systems
display the characteristics of Poisson distribution.
\par
To obtain the quantum mechanical energy levels of the system we can
write the hamiltonian (4) with the creation and destruction operators:
\beq
a_k=(r_k+i p_k),
\;\;\;\;
a_k^+ = (r_k -i p_k),
\eeq
where $k=1,2$, so we have:
\beq
H=\epsilon (a_1^+ a_1 + a_2^+ a_2 +1)-
{\chi \over 16} (a_1 +a_1^+)^2 (a_2 +a_2^+)^2.
\eeq
The eigenvalues can be calculated by diagonalizing the
hamiltonian in the basis $|n_1 n_2>$ of the occupation numbers of the two
harmonic oscillators. The matrix elements can be written:
$$
<n_{1}^{'}n_{2}^{'}|H|n_{1}n_{2}>=\epsilon (n_{1}+n_{2}+1)
\delta_{n_{1}^{'}n_{1}} \delta_{n_{2}n_{2}}+
$$
$$
+{\chi \over 16}
[\sqrt{n_{1}(n_{1}-1)} \delta_{n^{'}_{1}n_{1}-2}
+\sqrt{(n_{1}+1)(n_{1}+2)}\delta_{n^{'}_{1}n_{1}+2}+
(2n_{1}+1)\delta_{n^{'}_{1}n_{1}}]\times
$$
\beq
\times[\sqrt{n_2 (n_2-1)}\delta_{n^{'}_2 n_2-2}+ \sqrt{(n_2+1)(n_2+2)}
\delta_{n^{'}_2 n_2+2}+ (2n_2+1)\delta_{n^{'}_2 n_2}]
\eeq
and each submatrix can be
labelled by the parity of the occupation numbers $n_{1}$, $n_{2}$.
By performing the unfolding procedure described in detail in reference
[14], each spectrum has been mapped into one with quasi--uniform
level density.
\par
The distribution $P(S)$ of spacings between adjacent levels has been
calculated and compared to the Brody [15] distribution:
\beq
P(S)=\alpha (q+1) S^q exp{(-\alpha S^{q+1})},
\eeq
with:
\beq
\alpha = (\Gamma [{q+2\over q+1}])^{q+1}, \;\;\; 0\leq q\leq 1.
\eeq
The distribution interpolates between the Poisson distribution ($q=0$)
of integrable systems and the Wigner--Dyson distribution ($q=1$) of
chaotic ones. In figure 5 (a) the distribution $P(S)$ obtained with all
the four classes is plotted for $0\leq E\leq 12$ MeV with $q=0$,
and in (b) for $12 \leq E\leq 23$ MeV with $q=0.15$.

\vskip 0.5 truecm
{\bf 6. Conclusions}
\vskip 0.5 truecm
\par
In this paper we have shown that the quadrupole--quadrupole residual
interaction leads to the transition from order to chaos. However the
energy of the onset of chaos ($E_C\simeq 12.6$ MeV) is much higher
than the experimental one of atomic nuclei (chaotic behaviour $\simeq$ 8 MeV)
[5,6]. Therefore the quadrupole--quadrupole force is unable to reproduce
experimental nuclear data and higher multipole components
should be added [16].

\vskip 0.5 truecm
\begin{center}
{* * * * *}
\end{center}

\newpage

\parindent=0.pt

\section*{Figure Captions}
\vspace{0.6 cm}

Figure 1: The potential energy of the model ($A=200$).

Figure 2: The contour plot of the potential energy and the region of
negative curvature (marked by $\cdot$).

Figure 3: Ratio of stable points in phase space obtained from the
curvature criterion.

Figure 4: The Poincar\`e sections of the model:
(a) $E=7$ MeV, (b) $E=15$ MeV, (c) $E=23$ MeV.

Figure 5: The distribution $P(S)$ of spacings $S$ between adjacent
levels: (a) $0\leq E\leq 12$ MeV ($q=0$), (b) $12\leq E\leq 23$ MeV
(q=0.15).

\newpage

\section*{References}
\vspace{0.6 cm}

[1] O. Bohigas, H.A. Weidenm\"uller: Ann. Rev. Nucl. Part. Sci. {\bf
38}, 421 (1988)

[2] M.C. Gutzwiller: {\it Chaos in Classical and Quantum Physics}
(Springer-Verlag, 1991)

[3] {\it Chaos and Quantum Physics}, in Les Houches Summer School 1989, ed.
M.J. Giannoni, A. Voros, J. Zinn Justin (Elsevier Science Publishing,
1989); {\it From Classical to Quantum Chaos}, SIF Conference
Proceedings, vol. 41, Trieste 1992, ed. G.F. Dell'Antonio, S. Fantoni,
V.R. Manfredi (Editrice Compositori, 1993)

[4] B.R. Mottelson: in {\it Elementary Modes of Excitations in Nuclei},
Int. School of Phys. E. Fermi, Cours LXIX (1977)

[5] B.R. Mottelson: Nucl. Phys. A {\bf 557}, 717c (1993)

[6] M.T. Lopez--Arias, V.R. Manfredi, L. Salasnich: Riv. Nuovo Cimento
{\bf 17}, n.5, 1 (1994)

[7] M. Matsuo, T. Dossing, E. Vigezzi, R.A. Broglia:
Phys. Rev. Lett. {\bf 70}, 2694 (1993);
M. Matsuo, T. Dossing, B. Herskind, S. Frauendorf, E. Vigezzi, R.A.
Broglia: Nucl. Phys. A {\bf 557}, 211c (1993)

[8] R.D. Lawson: {\it Theory of the Nuclear Shell Model} (Oxford Science
Publications, 1980)

[9] M. Toda: Phys. Lett. A {\bf 48}, 335 (1974);
Yu.L. Bolotin, V.Yu. Gonchar, E.V. Inopin, V.V. Levenko,
V.N. Tarasov, N.A. Chekanov: Sov. J. Part. Nucl. {\bf 20}, 372 (1989)

[10] G. Benettin, R. Brambilla, L. Galgani: {\it Physica} A {\bf 87},
381 (1977)

[11] Subroutine D02BAF, The NAG Fortran Library, Mark 14, Oxford: NAG
Ltd. and USA: NAG Inc. (1990)

[12] H. Poincar\`e: {\it New Methods of Celestial Mechanics}, vol. 3, ch.
27 (Transl. NASA Washington DC 1967)

[13] M. Henon: Physica D {\bf 5}, 412 (1982)

[14] V.R. Manfredi: Lett. Nuovo Cimento {\bf 40}, 135 (1984)

[15] T.A. Brody: Lett. Nuovo Cimento {\bf 7}, 482 (1973)

[16] K. Arita, K. Matsuyanagi: Prog. Theor. Phys. {\bf 89}, 389 (1993);
W.D. Heiss, R.G. Nazmitdinov, S. Radu: Phys. Rev. Lett. {\bf 72},
2351 (1994)

[17] R. Badii: in {\it Chaotic Dynamics: Theory and Practice}, ed. T.
Bountis, NATO ASI Series B {\bf 298} (1992)

\end{document}